\documentclass[prd,twocolumn,showpacs,showkeys,superscriptaddress,floatfix,amsmath,nofootinbib,amssymb]{revtex4}

\usepackage{graphicx}
\usepackage{dcolumn}
\usepackage{mathrsfs}
\usepackage{color,xspace}

\newcommand{\uin}[1]{{u_{{#1}}^{\text{in}}}}
\newcommand{\uout}[1]{{u_{{#1}}^{\text{out}}}}
\newcommand{\uh}[1]{{u_{{#1}}^{{\text{hor}}}}}
\newcommand{\ain}[1]{{a_{{#1}}^{\text{in}}}}
\newcommand{\aout}[1]{{a_{{#1}}^{\text{out}}}}
\newcommand{\ket}[1]{\left| {#1} \right\rangle}
\newcommand{\bra}[1]{\left\langle {#1} \right|}
\newcommand{\ah}[1]{{a_{{#1}}^{{\text{hor}}}}}

\newcommand{\tr}{\operatorname{tr}}
\newcommand{\eo}{\eta_{\text{out}}}
\newcommand{\diff}{\text{d}}

\begin{document}

\title{Quantum entanglement produced in the formation of a black hole}
\author{Eduardo Mart\'{i}n-Mart\'{i}nez}
\affiliation{Instituto de F\'{i}sica Fundamental, CSIC, Serrano 113-B, 28006 Madrid, Spain}
\author{Luis J. Garay}
\affiliation{Departamento de F\'isica Te\'orica II, Universidad Complutense de Madrid, 28040 Madrid, Spain}
\affiliation{Instituto de Estructura de la Materia, CSIC, Serrano 121, 28006 Madrid, Spain}
\author{Juan Le\'on}
\affiliation{Instituto de F\'{i}sica Fundamental, CSIC, Serrano 113-B, 28006 Madrid, Spain}

\date{August 26, 2010}

\begin{abstract}
A field in the vacuum state, which is  in principle  separable, can evolve to an entangled state  in  a dynamical gravitational   collapse. We will study, quantify, and discuss the origin of this entanglement, showing that it could even reach the maximal entanglement limit for low frequencies or very small black holes, with consequences in micro-black hole formation and the final stages of evaporating black holes. This entanglement provides quantum information resources between the modes in the asymptotic future (thermal Hawking radiation) and those which fall to the event horizon. We will also show that   fermions are more sensitive than bosons to this quantum entanglement generation. This fact could be helpful in finding experimental evidence of the genuine quantum Hawking effect in analog models.
\end{abstract}



\maketitle

\section{Introduction}

Quantum entanglement has been recognized to play a key role in black
hole thermodynamics and the fate of information  in the presence of
horizons; some previous studies were performed in stationary cases,
namely the eternal acceleration scenario and the stationary eternal
Schwarzschild black hole \cite{Alsingtelep,Alicefalls,AlsingSchul,Edu2,Edu3,Edu4,Edu5,Edu6}, not addressing
issues related with dynamics and time evolution of gravitating quantum
fields. On the other hand some studies involving entanglement dynamics in
expanding universe  scenarios have shown that the interaction with the
gravitational field can produce entanglement between quantum field
modes \cite{Ball,Edu7}.

In this paper we analyze the issue of entanglement production in a
dynamical gravitational collapse. With this aim, we consider both a bosonic
(scalar) and a fermionic (Grassmann scalar) field which initially are  in the
vacuum state and compute their asymptotic time evolution under   the
gravitational interaction in a stellar collapse. The vacuum state  evolves to
an  entangled state of  modes in the future null infinity (which give rise to
Hawking radiation \cite{Hawking}) and modes that do not reach it
because they fall into the forming event horizon.

We will argue that the initial vacuum state  in the asymptotic past does
not have any physical quantum entanglement, and that it evolves to a state
that is physically entangled  as a consequence of the creation of the
event horizon. This entanglement depends on the mass of the black hole
and the frequency of the field modes. In particular, for very small
frequencies or very small black holes, a maximally entangled state could
be produced.

The entanglement generated in a gravitational collapse thus
appears as a quantum resource for non-demolition methods aiming to
extract information about the field modes which fall into the horizon
from the outgoing Hawking radiation. These methods would be most
relevant for cases such as the formation of micro-black holes and the
final stages of an evaporating black hole when the mass is getting smaller
and, therefore, quantum correlations generated between the Hawking
radiation and the infalling modes grow to become even maximal, as we will show.

Earlier works proved that fermions and bosons  have qualitatively
different behaviors in phenomena such as the Unruh entanglement
degradation \cite{Alicefalls,Edu4,Edu5}  and the entanglement
generation in the background of expanding universes \cite{Ball,Edu7}.
Here, we will show that for fermions the generation of entanglement due
to gravitational collapse is more robust than for bosons. This robustness
is more evident from the peak of the thermal spectrum of Hawking
radiation towards the ultraviolet.

Previous works in the literature (see for example
\cite{Balbinot,NavarroSalas,BalbinotII,Serenada} among many others)
showed that Hawking radiation is correlated with the field state falling
into the collapsing star. However neither the analysis of the associated
entanglement entropy as a function of the black hole parameters nor the
comparison between fermionic and bosonic behavior have been carried
out so far. The study of these issues,  the nature of the entanglement
produced in a gravitational collapse and, more important, its dependence
on the nature of the quantum field (bosonic/fermionic)  is decisive in
order to gain a deeper understanding about quantum entanglement in
general relativistic scenarios as it was proven for other setups such as
acceleration horizons, eternal black holes and expanding universes
\cite{Alicefalls,Edu4,Edu5,Ball,Edu7}.

Since entanglement is a pure quantum effect,  understanding its behavior
in these scenarios can well be relevant to discern the genuine quantum
Hawking radiation from classical induced emission in black hole analogs
\cite{Unruhan} (see, for example, Ref. \cite{serena}), where both
classical and quantum perturbations obey the same evolution laws.  It will
also follow from our study  that fermionic modes could be more suitable
for this task since they are more reliable in encoding entanglement
information.

Finally,  we will argue that the entanglement between the infalling and the
Hawking radiation modes neither existed as a quantum information
resource nor could have  been acknowledged by any observer before the
collapse occurs, namely in the asymptotic past. This is important in order
to understand the dynamics of the creation of correlations in the
gravitational collapse scenario since these correlations are exclusively
due to quantum entanglement, as discussed in the literature
\cite{Balbinot,NavarroSalas,BalbinotII,Serenada}.

\section{Gravitational Collapse}

\begin{figure}
\includegraphics[width=.9\columnwidth]{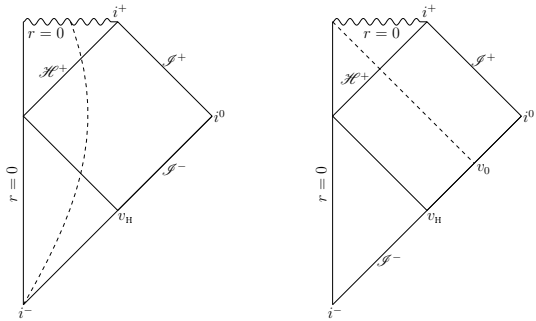}
\caption{Carter-Penrose diagrams for gravitational collapse: Stellar collapse (left)
and Vaidya spacetime (right).}
\label{fig:vaidya}
\end{figure}

In order to analyze the entanglement production induced by gravitational
collapse we will consider the Vaidya dynamical solution to
Einstein equations (see e.g. Ref. \cite{NavarroSalas})  that, despite its
simplicity, contains all the ingredients relevant to our study. Refinements
of the model to make it more realistic only introduce subleading
corrections. The Vaidya spacetime (Fig.~\ref{fig:vaidya}), is conveniently
described in terms of ingoing Eddington-Finkelstein coordinates by the
metric
\begin{equation}
\diff s^2=-\left(1-\frac{2M(v)}{r}\right)\diff v^2+2\diff v\diff r+r^2 \,
\diff \Omega^2,
\end{equation}
where $r$ is the radial coordinate, $v$ is the ingoing null coordinate, and
$M(v)=m\theta(v-v_0)$. For $v_0<v$ this is nothing but the ingoing
Eddington-Finkelstein representation for the Schwarzschild metric
whereas for $v<v_0$ it is just Minkowski spacetime. This metric
represents a radial  ingoing collapsing shockwave of radiation. As it can
be seen in Fig. \ref{fig:vaidya}, $v_\textsc{h}=v_0-4m$ represents the
last null ray that can escape to the future null infinity $\mathscr{I}^+$
and hence that will eventually form the event horizon.

Let us now introduce two different bases of solutions to the
Klein-Gordon equation in this collapsing spacetime. On the one hand, we
shall define the `in' Fock basis in terms of ingoing positive frequency
modes, associated with the natural time parameter $v$ at the null past
infinity $\mathscr{I}^-$, which is a Cauchy surface:
\begin{equation}
\uin{\omega}\sim\frac{1}{4\pi r\sqrt{\omega}}e^{-i\omega v}.
\end{equation}
On the other hand, we can also consider an alternative Cauchy surface in
the future to define another basis. In this case, the asymptotic future
$\mathscr{I}^+$ is not enough and we also need the future event horizon
$\mathscr{H}^+$. The `out' modes defined as being outgoing
positive-frequency in terms  of the natural time parameter
$\eta_{\text{out}}$ at  $\mathscr{I}^+$ are
\begin{equation}
\uout{\omega}\sim\frac{1}{4\pi r\sqrt{\omega}}e^{-i\omega \eta_{\text{out}}},
\end{equation}
where $\eo=v-2r^*_{\text{out}}$ and $r^*_\text{out}
$ is the radial tortoise coordinate in the Schwarzschild region.  It can be
shown (see e.g. Ref. \cite{NavarroSalas}) that, for late times
$\eo\to\infty$ at $\mathscr{I}^+$, these modes $\uout{\omega}$ are
concentrated near $v_\textsc{h}$ at $\mathscr{I}^-$ and have the
following behavior:
\begin{equation}\label{uout2}
\uout{\omega}\approx\frac{1}{4\pi r \sqrt{\omega}}
e^{-i\omega\left(v_\textsc{h}-4m\ln\frac{|v_\textsc{h}-v|}{4m}\right)}
\theta(v_\textsc{h}-v).
\end{equation}
These modes have only support in the region $v < v_\textsc{h}$. This is
evident as only the light rays that depart from $v < v_\textsc{h}$ will
reach the asymptotic region $\mathscr{I}^+$ since the rest will fall into
the forming horizon defined by $v=v_\textsc{h}$. This is the only
relevant regime, as far as entanglement production is concerned.

For the `hor' modes defined at  $\mathscr{H}^+$, there is no such
natural time parameter. A simple way to choose these modes is defining
them as the modes that in the asymptotic past $\mathscr{I}^-$ behave in
the same way as $\uout{\omega}$ but defined for $v > v_\textsc{h}$,
that is to say, as modes that leave the asymptotic past but do not reach
the asymptotic future  since they will fall into the horizon. This criterion
is the simplest that clearly shows the generation of quantum
entanglement between the field in the horizon and the asymptotic region.
In any case, since we will trace over all modes at the horizon, the choice
of such modes does not affect the result. Therefore, we define the
incoming modes crossing the horizon  by reversing the signs of
$v_\textsc{h}-v$ and $\omega$ in
\eqref{uout2} so that, near $\mathscr{I}^-$ these modes are
\begin{equation}
\uh{\omega}\sim\frac{1}{4\pi r \sqrt{\omega}}
e^{i\omega\left(v_\textsc{h}-4m\ln\frac{|v_\textsc{h}-v|}{4m}\right)}
\theta(v-v_\textsc{h}).
\end{equation}

We are now ready to write the annihilation operators of bosonic field
modes in the  asymptotic past   in terms of the corresponding creation
and annihilation operators defined in terms of modes in the future:
\begin{eqnarray}\label{ain}
\nonumber\ain{\omega'}&=&\int d\omega \Big[\alpha^*_{\omega\omega'}
\big(\aout{\omega}-\tanh r_{\omega}\,\ah{\omega}^\dagger\big)\\
&&+\alpha_{\omega\omega'}e^{i\varphi}\big(\ah{\omega}-\tanh r_{\omega}\,
\aout{\omega}^\dagger\big)\Big],
\end{eqnarray}
where  $\tanh r_{\omega} = e^{-4\pi m\omega}$. The precise
values of $\varphi$ and  $\alpha_{\omega\omega'}$ are not relevant for
this analysis.

Hence the vacuum $\ket{0}_\text{in}$, annihilated by \eqref{ain} for all
frequencies $\omega'$, acquires the following form in terms of the
`out-hor' basis:
\begin{equation}
\ket{0}_\text{in}=N \exp\left(\sum_{\omega}\tanh r_{\omega}\,
\ah{\omega}^\dagger \aout{\omega}^\dagger\right)
\ket{0}_{\text{hor}}\ket{0}_{\text{out}},
\end{equation}
where $N=\big(\prod_{\omega}\cosh r_{\omega}\big)^{-1}$ is a
normalization constant. We can rewrite this state in terms of modes
$\ket{n_{\omega}}$ with frequency $\omega$ and occupation number
$n$ as
\begin{equation}\label{sque}
\ket{0}_\text{in}=\prod_{\omega}\frac{1}{\cosh r_{\omega}}
\sum_{n=0}^\infty (\tanh r_{\omega})^{n}\ket{n_{\omega}}_\text{hor}
\ket{n_{\omega}}_{\text{out}}.
\end{equation}

\section{Analyzing entanglement}

This is a two-mode squeezed state.  Therefore, it is a pure  entangled
state of the modes   in the asymptotic future and the modes falling
across  the event horizon. Given the tensor product structure no
entanglement is created between different frequency modes. Hence, we
will concentrate the analysis in one single arbitrary frequency $\omega$.

\begin{figure}
\begin{center}
\includegraphics[width=.90\columnwidth,height=.5\columnwidth]{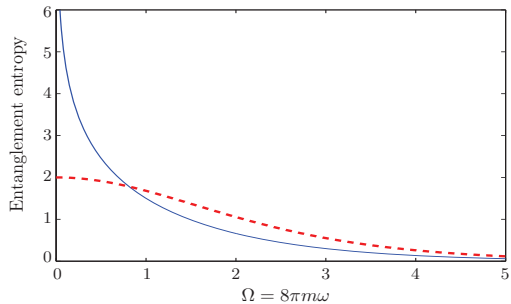}
\caption{Entanglement between bosonic (continuous blue) and fermionic (red dashed)
field modes in $\mathscr{H}^+$ and in $\mathscr{I}^+$.
The lesser the mass of the star or the mode frequency, the higher the
entanglement reached.}
\label{fig2}
\end{center}
\end{figure}

We can compute the entropy of entanglement for this state which is the
ultimate entanglement measure for a bipartite pure state,
defined as the Von Neumann entropy of  the reduced state obtained upon
tracing over one of the subsystems of the bipartite state. To compute it
we need the partial state
$\rho_{\text{out}}=\tr_{\text{hor}}(\ket{0}_{\text{in}}\!\!\bra{0})$,
which turns out to be
$\rho_{\text{out}}=\prod_\omega\rho_{\text{out},\omega}$, where
\begin{equation}
\rho_{\text{out},\omega}=\frac{1}{(\cosh r_{\omega})^2}
\sum_{n=0}^\infty (\tanh r_{\omega})^{2n}
\ket{n_{\omega}}_{\text{out}}\!\!\bra{n_{\omega}}.
\end{equation}
This is, indeed, a thermal radiation state  whose temperature is nothing
but the Hawking temperature $(8\pi m)^{-1}$, as  it can be easily
seen. However, this is only the partial state of the field, not the complete
quantum state, which is globally  entangled. If we compute the entropy of
entanglement
$S_{\textsc{e},\omega}=\tr(\rho_{\text{out},\omega}\log_2
\rho_{\text{out},\omega})$ for each frequency, after some calculations, we obtain
\begin{align}
\nonumber S_{\textsc{e},\omega}&=
\left(\cosh r_\omega\right)^2\log_2\left(\cosh r_\omega\right)^2
\\&-(\sinh r_\omega)^2\log_2\left(\sinh r_\omega\right)^2,
\end{align}
which is displayed in Fig.~\ref{fig2}. As \eqref{sque} is a pure state, all
the correlations between modes at the horizon and modes in the
asymptotic region are due to quantum entanglement.

Analogously we can compute the entanglement for fermionic
fields. If we consider a spinless Dirac field (either one dimensional or a
Grassmann scalar), the analysis is entirely analogous considering
now both particle and antiparticle modes. We assume again that the
initial state of the field is the vacuum that, after some long nontrivial
calculations, can be expressed it in the Fock basis at the asymptotic
future and the `hor' modes:
\begin{align}\label{squef}
\ket{0}_\text{in}&=\prod_{\omega}\Big[(\cos\tilde r_{\omega})^2
\ket{00}_\text{hor}\ket{00}_{\text{out}}
\nonumber\\
&-\frac{\sin 2\tilde r_{\omega}}{2}
\big(\ket{01_\omega}_\text{hor}\ket{1_\omega0}_{\text{out}}
 - \ket{1_\omega0}_\text{hor}
 \ket{01_\omega}_{\text{out}}\big)
 \nonumber\\
 &-(\sin\tilde r_{\omega})^2
 \ket{1_\omega1_\omega}_\text{hor}
 \ket{1_\omega1_\omega}_{\text{out}}\Big],
\end{align}
where  $\tan \tilde r_\omega= e^{-4\pi m\omega} $. Here, we are using
the double Fock basis, the first figure inside each ket representing
particles and the second  antiparticles.

We can compute the entropy of entanglement of this pure state.
The partial density matrix in the asymptotic future
$\rho_{\text{out}}=\tr_{\text{hor}}(\ket{0}_{\text{in}}\!\!\bra{0})=
\prod_\omega \rho_{\text{out},\omega}$, is given by
\begin{align}
\rho_{\text{out},\omega}&= (\cos\tilde r_\omega)^4
\ket{00}_{\text{out}}\!\!\bra{00}
\nonumber\\
&+\frac{(\sin 2\tilde r_\omega)^2}{4}
\Big(\ket{1_{\omega}0}_{\text{out}}\!\!\bra{1_{\omega}0}
+\ket{01_{\omega}}_{\text{out}}\!\!\bra{01_{\omega}}\Big)
\nonumber\\
&+(\sin \tilde r_\omega)^4 \ket{1_{\omega}1_{\omega}}_{\text{out}}\!\!
\bra{1_{\omega}1_{\omega}},
\end{align}
which   is again a thermal state with Hawking temperature $(8\pi
m)^{-1}$, and
\begin{equation}
S_{\textsc{e},\omega}\!=\!-2\big[(\cos \tilde r_\omega)^2\log_2
(\cos \tilde r_\omega)^2+
(\sin \tilde r_\omega)^2\log_2(\sin \tilde r_\omega)^2\big],
\end{equation}
which is also displayed in Fig.~\ref{fig2}.

Figure~\ref{fig2} shows that the entanglement decreases as the mass of
the black hole or the frequency of the mode increase. When comparing
bosons with  fermions one must have in mind that the entropy of
entanglement is bounded by (the logarithm of) the dimension of the
partial Hilbert space (`out' Fock space in our case). Therefore, due to
Pauli exclusion principle, the maximum entropy of entanglement for
fermions is $S_{\textsc{e},\omega}=2$, which corresponds to a
maximally entangled state. On the other hand, for bosons, the
entanglement is distributed among the superposition of all the occupation
numbers and the entropy can grow unboundedly, reaching the
maximally entangled state in the limit of infinite entropy. In this sense,
the entanglement generated in the fermionic case is more useful and
robust due to the limited dimension of the Fock space for each fermionic
mode.

This result can be traced back to the inherent differences between
fermions and bosons. Specifically, it is Pauli exclusion principle which
makes fermionic entanglement more reliable.  Similar results about
realiability of entanglement for fermions were also found in the
expanding universe scenarios \cite{Edu7}.  This responds to the high
influence of statistics in entanglement behaviour in general relativistic
settings as it was investigated in \cite{Edu4,Edu5}. On the other hand,
Vaidya space-time has all the fundamental features of a stellar collapse
and shows how the entanglement is created by the appearance of an
event horizon. Hence, in other collapsing scenarios or including the
sub-leading grey-body factor corrections, these fundamental statistical
differences will not disappear. The qualitatively different behavior of
entanglement for bosons and fermions is not an artifact of choosing a
particular collapse scenario but is due to fundamental statistical
principles.

In the above analysis we have considered plane wave modes, which are
completely delocalized. However, an entirely analogous analysis can be easily
carried out using very well localized Gaussian states, with the same
results about quantum entanglement behavior.

\section{Conclusions}

We have shown that the formation of an event horizon generates
entanglement. If we start from the vacuum state in the asymptotic past,
after the gravitational collapse process is complete we end up with a
state in the asymptotic future which shares pure quantum correlations
with the field modes which fall into the horizon. One could think that this
entanglement was already present before the   collapse, arguing that (as
proved in \cite{Vacbell}) the vacuum state of a quantum field can be
understood as an entangled state of space-like separated regions. In
other words, if we artificially divided  the Cauchy surface in which the
vacuum state is determined into two parts, we would have a quantum
correlated state between the two partitions. In principle  we could have
done a bipartition of the vacuum state in $\mathscr{I}^-$ such that it
would reflect entanglement between the partial state of the vacuum for
$v<v_\textsc{h}$ and the corresponding partial state for
$v>v_\textsc{h}$. However, it is not until the collapse occurs that we
have the information about what $v_\textsc{h}$ is. So, achieving
beforehand the right bipartition (trying to argue that the entanglement
was already in the vacuum state) would require a complete knowledge of
the whole future and, consequently, there is no reason `a priori' to do
such bipartition. The entanglement, eventually generated by the collapse,
will remain unnoticed to early observers, who are deprived of any means
to acknowledge and use it for quantum information tasks. It is well known
that if we introduce artificial bipartitions of a quantum system, its
description can show entanglement as a consequence of the partition.
However, not being associated with a physical bipartition this
entanglement does not codify any physical information and, hence, cannot
be used to perform any quantum information processes. (One example of
this kind of non-useful entanglement is statistical entanglement between
two undistinguishable fermions~\cite{sta1}).

Gravitational collapse selects a specific partition  of the initial vacuum
state by means of the creation of an event horizon. In the asymptotic
past there was no reason to consider a specific bipartition of the vacuum
state, whereas in the future there is a clear physically meaningful
bipartition: What in $\mathscr{I}^-$ was expressed as a separable state,
now becomes expressed in terms of modes that correspond to the future
null infinity and the ones which fall across the event horizon. This means
that gravitational collapse defines a particular physical way to break the
arbitrariness of bipartitioning the vacuum into different  subsystems.
This gravitational production of entanglement would be a physical
realization  of the potentiality of the vacuum state to be an entangled
state and is therefore a genuine entanglement creation process.

We have computed the explicit functional form of this entanglement and
its dependence on the mass of the black hole (which determines the
surface gravity). For more complicated scenarios (with charge or angular
momentum), it will depend on these parameters as well.

For small black holes, the outgoing Hawking radiation tends to be
maximally entangled with the state of the field falling into the horizon
for both bosons and fermions. This means that if a hypothetical high
energy process generates a micro-black hole, a projective measurement
carried out on the emitted radiation (as, for instance, the detection of
Hawking radiation) will `collapse' the quantum state of the field that is
falling into the event horizon and give us certainty about the outcome of
possible measurements carried out in the vicinity of the horizon.
Furthermore, at least theoretically speaking, the available quantum
information resources would be maximum and, therefore, one could
perform quantum information tasks such as  quantum teleportation with
maximum fidelity from the infalling modes to the modes in the asymptotic
future $\mathscr{I}^+$ if the observer of the infalling modes managed
to dispatch an outgoing classical signal  before crossing the horizon. On
the other hand, low frequency modes become more entangled than the
higher ones. So, the infrared part of the Hawking spectrum would provide
more information about the state at the horizon than the ultraviolet.

Arguably, similar conclusions can be drawn for the final stages of an
evaporating black hole: As the mass of the black hole diminishes, the
temperature of the Hawking radiation spectrum increases, and
therefore, the quantum state of the field tends to a maximally entangled
one in the limit of $m\rightarrow0$.

We have seen that the entanglement generated in fermionic fields is more
robust than for bosons. Although the entropy of entanglement in the zero
mass limit is greater in the bosonic case due to the higher dimension of
the partial Hilbert space, we have argued that the information is more
reliably encoded in the limited Fock space of fermionic fields.
Furthermore, as we consider higher frequency modes, fermionic
entanglement proves to be much more   easily created by the collapse.
What is more, the turning point in which the entropy of entanglement for
fermions becomes numerically larger than for bosons is actually near the
peak of the thermal emission (Fig.~\ref{fig2}). This means, that, in
general, a measurement carried out on Hawking radiation of fermionic
particles will give us more information about the near-horizon field state.
This might also be useful in analog gravity realizations as we have already
discussed,  specifically in systems where the field excitations are
fermionic (see e.g. Ref. \cite{Volovik}), which would be, as shown, at an
advantage over the bosonic cases. To account for this quantum
entanglement in analog experiments one should  carry out measurements
of the quantum correlations between the emitted thermal spectrum and
the infalling modes and detect Bell inequalities violations. This is easier as
it gets closer to the maximally entangled case.

\section{Acknowledgements}

The authors want to thank Carlos Barcel\'{o} for useful discussions. This
work was  supported by the Spanish MICINN Projects
FIS2008-05705/FIS, FIS2008-06078-C03-03, the CAM research
consortium QUITEMAD S2009/ESP-1594, and the Consolider-Ingenio
2010 Program CPAN (CSD2007-00042). E. M-M was partially supported
by a CSIC JAE-PREDOC2007 grant.

\bibliographystyle{apsrev}

\end{document}